\documentclass[%
 aip,
 amsmath,amssymb,
jcp,%
]{revtex4-2}

\usepackage{xcolor}
\usepackage{multirow}
\usepackage{upgreek}
\usepackage{graphicx}
\usepackage{geometry}
\usepackage{hyperref}
\usepackage{soul,color}

\begin{document}


\title{Ephemeral Ice-Like Local Environments in Classical Rigid Models of Liquid Water} 



\author{Riccardo Capelli}
\email{riccardo.capelli@polito.it}
\affiliation{Department of Applied Science and Technology, Politecnico di Torino, Corso Duca degli Abruzzi 24, I-10129 Torino, Italy}
\author{Francesco Muniz-Miranda}
\affiliation{Department of Chemical and Geological Sciences, University of Modena and Reggio-Emilia, Via Campi 103, I-41125 Modena, Italy}
\author{Giovanni M. Pavan}
\email{giovanni.pavan@polito.it}
\affiliation{Department of Applied Science and Technology, Politecnico di Torino, Corso Duca degli Abruzzi 24, I-10129 Torino, Italy}
\affiliation{Department of Innovative Technologies, University of Applied Sciences and Arts of Southern Switzerland, Polo Universitario Lugano, Campus Est, Via la Santa 1, CH-6962 Lugano-Viganello, Switzerland}



\date{\today}

\begin{abstract}
Despite great efforts over the past 50 years, the simulation of water still presents significant challenges and open questions. 
At room temperature and pressure, the collective molecular interactions and dynamics of water molecules may form local structural arrangements that are non-trivial to classify. 
Here we employ a data-driven approach built on Smooth Overlap of Atomic Position (SOAP) that allows us to compare and classify how widely used classical models represent liquid water. 
Macroscopically, the obtained results are rationalized based on water thermodynamic observables. 
Microscopically, we directly observed how transient ice-like ordered environments may dynamically/statistically form in liquid water, even above the freezing temperature, by comparing the SOAP spectra for different ice structures with those of the simulated liquid systems. 
This confirms recent \textit{ab initio}-based calculations, but also reveals how the emergence of ephemeral local ice-like environments in liquid water at room conditions can be captured by classical water models.\end{abstract}


\maketitle 

\section{Introduction}
Water is one of the most ubiquitous substances on the Earth's surface. Yet, a complete understanding of its structural properties, fundamental for its role in Life\cite{frappart,hassanali2018.3}, remains out of reach. 
In fact, the molecular arrangements observed in different phases of water are extraordinarily various and complex, resulting in a plethora of unusual and often not yet fully understood properties.\cite{Ball2008,gallo2016water,chaplinsite} 
From a microscopic point of view, such a complexity arises from a concerted action of directional and strong hydrogen bonds and non-specific Van der Waals interactions, which are in dynamic competition with the collective motion of the water molecules at finite temperature regimes.
Traditionally, the minimal picture to interpret the 1\textsuperscript{st} solvation shell shows each water molecule in condensed phase engaged in 4 hydrogen bonds with its immediate neighbors, yet in the last decades even this basic understanding has been challenged.\cite{Wernet995,huang2009inhomogeneous,nilsson2015structural} 
New interpretations have been proposed to explain water's less-common phases\cite{stanley1992,fanetti2014,hassanali2018.2}, showing how its detailed phase diagram is still subject to intense debate, both from an experimental\cite{niskanen2019compatibility,salzmann2019} and computational perspective.\cite{pipolo2017,cheng2021}

Indeed, achieving a satisfactory accuracy in the modeling of liquid water is a major challenge.
Simulation approaches at a high level of theory, \textit{e.g.}, those treating the electronic structure at the quantum-mechanical level (usually in its density-functional theory approximation) can accurately describe the covalent interactions\cite{Hassanali20410,Hassanali13723,hassanali2014} and hydrogen-bond dynamics\cite{muniz2012,gasparotto2016} of water molecules. Experimental properties such as the O$\dots$H radial distribution function and other observed behaviors\cite{ceriotti2013pnas,machida2018,gibbo2014,kuhne2015} also require further corrections to account for nuclear quantum effects to be perfectly matched.
In general, water has the tendency to display large local fluctuations\cite{sosso2017,hassanali2018}, a fact that has consequences in phenomena such as supercooling\cite{fitzner2009ice,hassanali2020,HU2022118334} or in the emergence of many metastable ``phases''.\cite{stanley1992,Handle13336}
Nonetheless, the high computational cost associated to these accurate quantum approaches to describe such phenomena typically hinders the possibility to achieve the ergodic limit for large samples of liquid water systems. 

The need of a reasonable compromise between computational cost and accuracy of the results (especially for biological applications) paved the way to the design of many classical force field potentials. 
Differently from the \textit{ab initio} approach, standard classical potentials lack polarization and treat molecules as rigid bodies to allow for longer time integration steps. 
With this kind of limitations in the system representation, the model performances are assessed on the comparison with known specific properties (obtained from experiments or from higher level of theory calculations).
In general, classical models for water would be expected to be accurate in the treatment of the liquid phase at room conditions, but sensible differences between models can still be observed and their development/optimization is still a very active research field. 

\begin{figure}[htbp]
\centering\includegraphics[width=\textwidth]{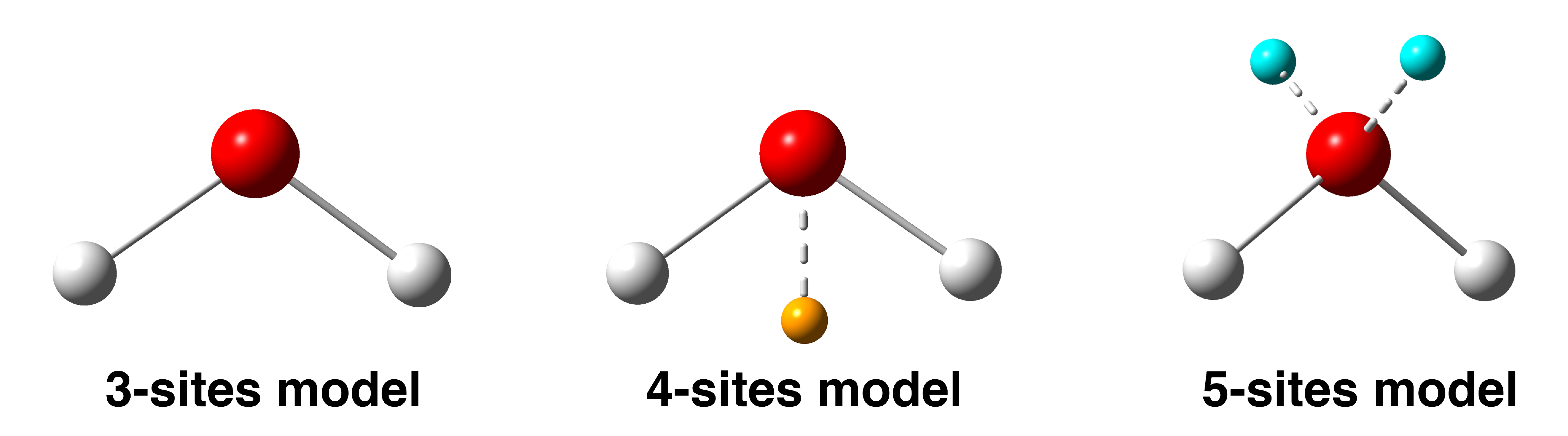}
\caption{Rigid classical models for the water molecule, including 3 point-charges (left), 4 point-charges (center), and 5 point-charges (right). Charges not coinciding with atomic positions are displayed in orange (center) and cyan (right).}
\label{fig:fig1}
\end{figure}

Historically, from the first simpler and still widely used 3-sites rigid water models (\textit{i.e.}, one site per atomic nucleus) developed by Berendsen \textit{et al.}\cite{berendsen1981} and by Jorgensen \textit{et al.}\cite{jorgensen1983}, the water models evolved to more complex representations, including one, two or even more virtual sites (see Figure \ref{fig:fig1} and Table \ref{tab:models}) to have an accurate accounting of effects that cannot be obtained \textit{via} classical potentials acting on only 3 nuclei. 
Indeed, the necessity of at least one virtual site in classical description of water for a qualitative reproduction of the phase diagram has been extensively demonstrated in previous works.\cite{vega2005can,abascal2007dipole}

Differences in the features of the water models usually may lead to large discrepancies in both the collective behavior and in the measured properties of water. 
But a direct correlation between microscopic parameters and macroscopic observables is not always straightforward in the simulation of systems containing a large number of explicit molecules. 
For this reason, the detection of ordered arrangements in liquid water models has been attempted using, \textit{e.g.}, order parameters\cite{errington2001relationship}, graph theory\cite{hassanali2018}, or environment descriptors\cite{monserrat2020liquid}. 
In an interesting recent study, using a machine learning-based potential of water trained on \textit{ab initio} calculations, Monserrat \textit{et al.} employed a local environment analysis which highlighted the presence of structures described as ``building blocks of ice phases'' in liquid water.\cite{monserrat2020liquid} 
A relevant question is whether similar ice-like precursor motifs can be encountered in liquid water also by classical water models, especially considered the importance of this in understanding the behavior of water and the central role played by classical models in nowadays molecular simulations.

Here we apply a data-driven approach based on Smooth Overlap of Atomic Position (SOAP) distances, which allow us to compare molecular systems and how different force fields treat them: some of us previously adopted it to classify lipid systems and classical lipid force fields\cite{capelli2021data} and to track defects and dynamic molecular arrangements in complex supramolecular systems\cite{gasparotto2019identifying,bian2021}. 
In the present paper, we employ this approach to compare the local fluctuating molecular environments that are formed in molecular dynamics simulations of liquid water, and to classify them based on their SOAP similarity with known ice environments.  
Macroscopically, we compared 16 of the most widely used rigid water models in liquid state and we classified how these represent the liquid water systems, correlating the obtained results with average thermodynamic observables of water. 
Microscopically, by comparing the SOAP spectra of geometrically-derived ice structures with those detected in the liquid systems during the MD simulations, we obtained evidence of the presence of local ice structures that may form and disappear during the MD even at temperature above the freezing one. In particular, we found presence of transient ordered ice-like structures in all the water models studied in this paper, which suggests that, at the very least, this is a common feature in the modeling itself of liquid water.

\section{Results and Discussion}

We started our analysis performing a series of 10 ns-long classical molecular dynamics (MD) simulations in the $NpT$ ensemble (with the three constants $NpT$ representing the number of particles, the pressure, and the temperature during the MD run, respectively) of identical cubic boxes filled with 1,024 water molecules for all the 16 rigid water models listed in Table \ref{tab:models}. 
For the best comparability, the simulation conditions were kept the same for all studied models (see the Methods section for complete computational details). 

\begin{table}[htbp]
\caption{Summary of the classical rigid water models compared herein.\\}
\label{tab:models}
\centering \small
\begin{tabular}{lccc} 
   & & & \\
   Model name  & Refs. & Year & Number of sites \\
   \hline
   SPC                              & ref.\cite{berendsen1981}          & 1981 & \multirow{6}{*}{3 Sites} \\
   SPC/E                            & ref.\cite{berendsen1987}          & 1987 & \\
   SPC/E\textsubscript{b}           & ref.\cite{takemura2012}           & 2012 & \\
   TIP3P                            & ref.\cite{jorgensen1983}          & 1983 & \\
   TIP3P-FB                         & ref.\cite{wang2014}               & 2014 & \\
   OPC3                             & ref.\cite{izadi2016accuracy}      & 2016 & \\  
   \hline
   TIP4P                            & ref.\cite{jorgensen1983} & 1983 & \multirow{7}{*}{4 Sites} \\
   TIP4P-Ew                         & ref.\cite{horn2004development}    & 2004 & \\
   TIP4P-FB                         & ref.\cite{wang2014}               & 2014 & \\
   TIP4P-ICE                        & ref.\cite{abascal2005potential}   & 2005 & \\
   TIP4P/2005                       &  ref.\cite{abascal2005general}     & 2005 & \\
   TIP4P/$\upvarepsilon$            &  ref.\cite{fuentes2014non}         & 2014 & \\
   OPC                              &  ref.\cite{izadi2014}              & 2014 & \\  
   \hline
   TIP5P                            &  ref.\cite{mahoney2000five}        & 2000 & \multirow{3}{*}{5 Sites}\\
   TIP5P-E                          &  ref.\cite{rick2004reoptimization} & 2004 & \\
   TIP5P/2018                       &  ref.\cite{khalak2018improved}     & 2018 & \\
   \hline
   \end{tabular}
\end{table}

\subsection{Macroscopic comparison with the experimental data}

\begin{figure}[htbp]
\centering\includegraphics[width=0.95\textwidth]{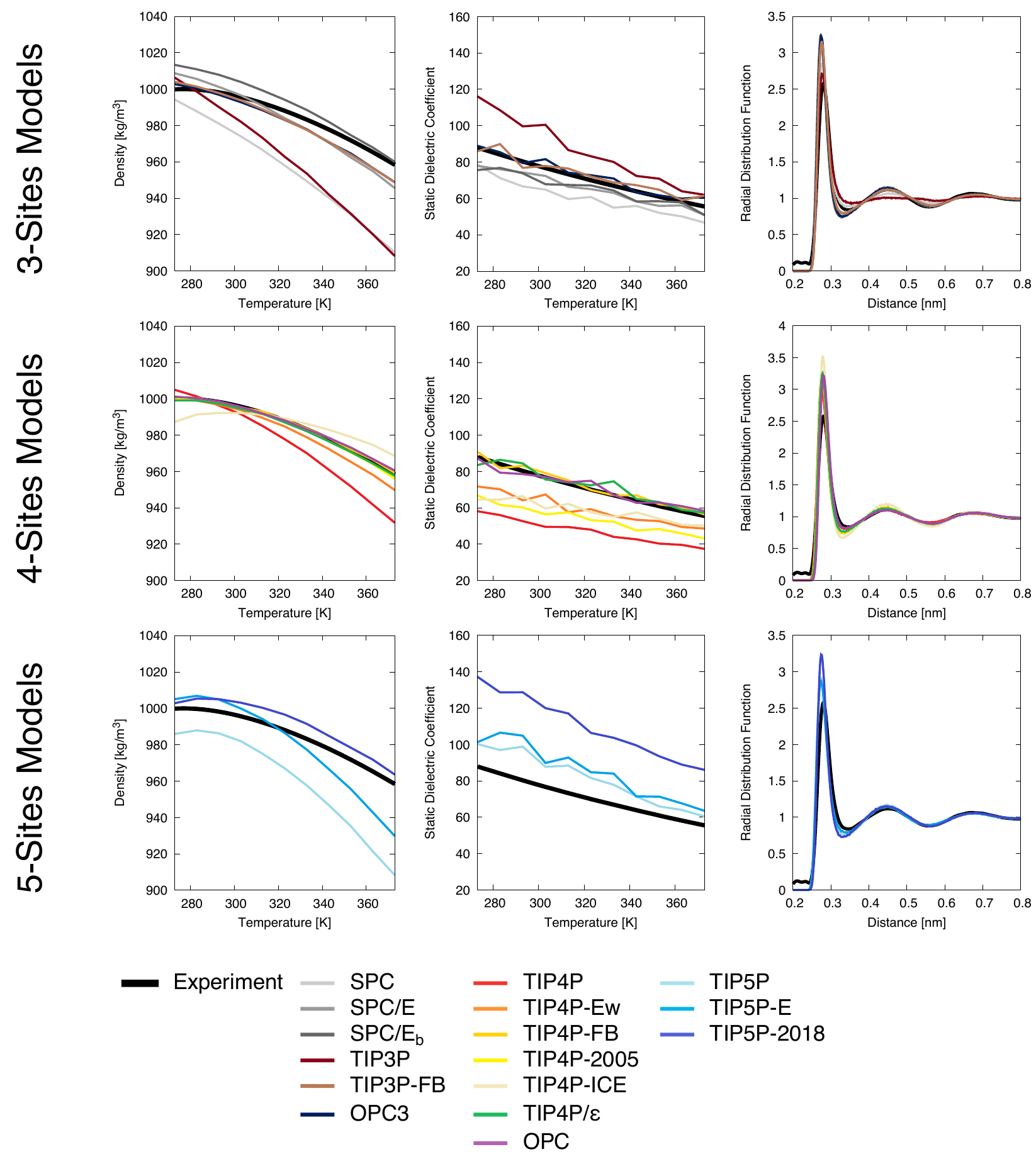}
\caption{Observables obtained from our simulations compared with experimental data. From left to right, liquid water density in function of temperature (experimental data from ref.\cite{haynes2014crc}), static dielectric constant in function of temperature (experimental data from ref.\cite{fernandez1997formulation}), and the radial distribution function at 298 K computed on oxygen atoms (experimental data from ref.\cite{skinner2013benchmark}). To have a fair comparison between all the models, the first row contains the models that have a 3-sites representation, then in the middle row the 4-sites representation, and in the last row we have the 5-sites models.}
\label{fig:fig2}
\end{figure}

From the equilibrium MD trajectories of all simulated systems, we computed a series of observables (displayed in Figure \ref{fig:fig2}): (1) the liquid water density, (2) the static dielectric constant, and (3) the pair radial distribution function of the oxygen atoms. 
Both the water density (1) and the dielectric constant (2) were calculated reported as functions of the temperature (Figure \ref{fig:fig2}: left and center), while the oxygen atoms' radial distribution functions were calculated at 298 K (Figure \ref{fig:fig2}: right). 

From such data, we can observe that an experimentally-consistent representation of the water density is out of reach for 3-sites models of water (Figure \ref{fig:fig2}: top-left panel). This is true also for the most recent models like TIP3P-FB and OPC3 (as it was also anticipated in the original publications)\cite{wang2014,izadi2016accuracy}. 
On the other hand, we observe an overall qualitative agreement with experiments for the density and dielectric constant trends for some 4-sites models (\textit{e.g.}, TIP4P/$\upvarepsilon$ and OPC), while the other 4-sites and, in general, all 5-sites models display poor agreement with the experimental observables.
All of them (with the exception of TIP4P and TIP4P/ICE) show a maximum around $\sim$277 K (Figure \ref{fig:fig2}: middle- and bottom-left panels), as expected from experiments.

Regarding the static dielectric constant ($\varepsilon$), we observe that a higher/lower number of sites is not crucial to have a satisfactory representation of it. 
Shown in Figure \ref{fig:fig2} (center), two models with 3-sites and three models with 4-sites representation display a good agreement with the experimentally observed $\varepsilon$. 
While this is somewhat expected for TIP4P/$\upvarepsilon$, which indeed has been parametrized with this main goal, this is a characteristic that is also well reproduced in recent parametrizations obtained by Wang \textit{et al.} (TIP3P-FB and TIP4P-FB),\cite{wang2014} and Izadi \textit{et al.} (OPC3 and OPC).\cite{izadi2014,izadi2016accuracy}

Regarding oxygen-oxygen radial distribution function (Figure \ref{fig:fig2}: right panels), we observe:
\begin{itemize}
    \item  a general discrepancy between all models and the experiments in terms of height of the 1\textsuperscript{st} radial distribution function peak, which has been generally imputed to the fact that quantum effects are not taken into account in such classical representations\cite{ceriotti2013pnas};
    \item  overall, a good representation of the subsequent solvation shells (up to at least the 3\textsuperscript{rd}) by all water models, with the notable exception of SPC and TIP3P models.  
\end{itemize}

From such first macroscopic analysis, the considered observables show that all the models and representations analyzed here are more or less comparable in terms of accuracy in the treatment of the bulk properties of liquid water. 
The largest deviation from the experimental data, in this sense, is observed for the oldest, yet widely employed, SPC and TIP3P models. 

\subsection{Microscopic SOAP analysis of liquid water models}
We opted for a data-driven analysis approach in order to gain a deeper insight into the differences/similarities between the various considered models and their representation of the bulk properties of water. 
Inspired by a recent work on the comparability of lipid force fields,\cite{capelli2021data} our approach uses a local environment descriptor, \textit{i.e.}, the Smooth Overlap of Atomic Position (SOAP),\cite{bartok2013,de2016comparing} and SOAP-based distances that allow us to compare and classify the various modeled systems. 
This SOAP-based analysis allows us to characterize the systems based on the molecular environments that are formed during the equilibrium MD simulation. 
In particular, centering one SOAP vector on the oxygen atom of every water molecule, this analysis substantially accounts for the geometrical arrangements of the molecules (\textit{i.e.}, both hydrogen and oxygen atoms) that surround each water molecule during MD simulations. 
In this way, we can compare the simulated systems based on the molecular environments that emerge in the systems themselves during the equilibrium MD simulations. 

\begin{figure}[htbp]
\centering\includegraphics[width=\textwidth]{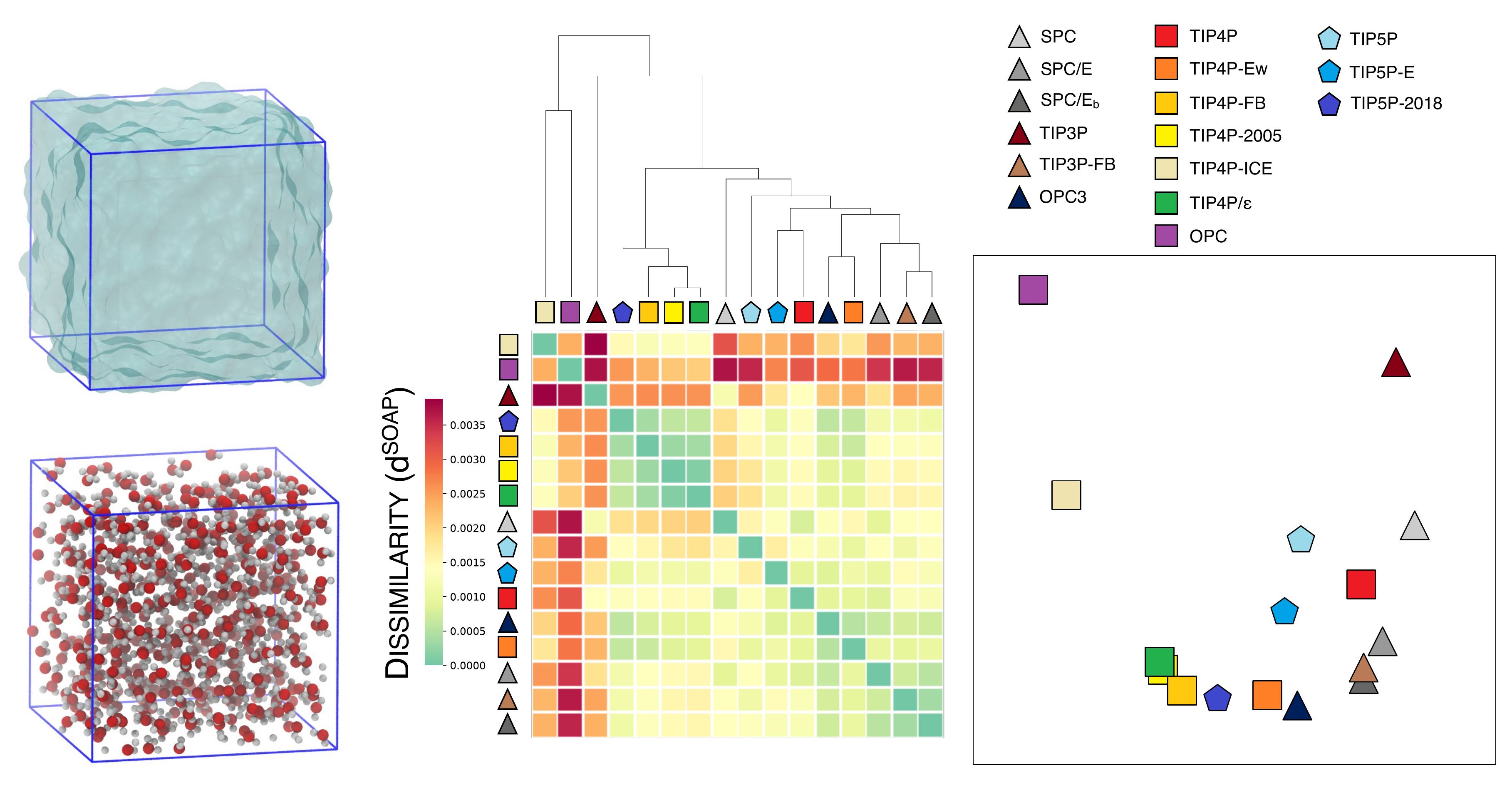}
\caption{SOAP distance analysis of different rigid water models. On the left we represented the cubic box at atomic detail (bottom right) and in ``liquid'' representation (top right). In the central panel we have a reciprocal SOAP distance matrix for all the water models analyzed; in the right panel, we have the MDS projection on a 2D plane of the distance matrix. The point shape is related to the number of sites of the model: triangle for 3-sites models, square for 4-sites, and pentagon for 5-sites representations.}
\label{fig:fig3}
\end{figure}

For this SOAP analysis, we considered all the simulations performed at $T$=298 K: the underlying assumption is that models parametrized to represent liquid water should work at their best at room conditions. 
As a first step, we computed the SOAP power spectra centered on the oxygen atoms of each water molecule in each system (1,024 water molecules in the simulation boxes). 
This has been done for each frame during the equilibrium MD trajectories (for a total of 1,001 MD frames analysed for each simulated model). The SOAP power spectra were computed using a cutoff of 1 nm and $l_{\text{max}}=n_{\text{max}}=8$ (see Methods section for complete details). The same parameters were used for the SOAP calculations of all systems studied in this work. 
Such a cutoff distance guarantees that the SOAP analysis takes into account at least up to the 3rd solvation shell around each water molecule.
Also, it has been recently shown that in liquid and liquid-like environments a larger cutoff is preferable rather than shorter ones.\cite{gasparotto2019identifying,capelli2021data}
In this way, we first obtain a dataset containing a total of 1,025,024 spectra (1,024 for each water molecule in the system, for each of the 1,001 sampled MD frames), that we then average, obtaining a final characteristic average SOAP spectrum for each simulated system. 
This analysis captures with high sensitivity all local environments present in the simulated systems, and collect them into a global high-dimensional SOAP spectrum that contains a unique \textit{multi-dimensional fingerprint}.
Similar to what recently done to classify lipid assemblies,\cite{capelli2021data} from the average SOAP spectra associated to each modelled system, we then computed a pairwise distance matrix, based on the metrics $d^{\text{SOAP}}$. 
Such $d^{\text{SOAP}}$ quantifies the difference between two spectra, measuring the distance between the simulated systems in the high-dimensional SOAP space (see Method section for additional details). This distance represents then their similarity/dissimilarity in terms of the local molecular environments formed in the liquid systems at equilibrium conditions.
The matrix we obtained is shown in Figure \ref{fig:fig3}, left panel. To provide a more informative representation of the data, we also projected the distance matrix on a 2D plane making use of the Multidimensional Scaling Algorithm (MDS)\cite{kruskal1964multidimensional}, reported in Figure \ref{fig:fig3}, right panel.  

From the scale in Figure \ref{fig:fig3}, it appears evident that the SOAP distances involved are very small, with magnitudes between the $\sim$ $10^{-4}-10^{-3}$ range. 
For sake of clarity, we underline here that $d^{\text{SOAP}}$ is by definition in the $[0,\sqrt{2}]$ interval, and the range of distances obtained between different force fields in a previous work on the comparison between lipid bilayers\cite{capelli2021data} gave $d^{\text{SOAP}}$ values in the order of $\sim$ $10^{-1}$. Furthermore, we analyzed the distribution of $d^{\text{SOAP}}$ among different environments in liquid water (Figure SI3), confirming that they spread in the SOAP space (as shown in Montserrat et al.\cite{monserrat2020liquid}).  
This observation is reassuring, because it means that, even on a more microscopic level, the many classical models analyzed here yield globally similar bulk water average structures. 
Yet, it is worth noting that the $d^{\text{SOAP}}$ values, albeit small, are non-negligible, and indeed some models are more/less distant that other ones.

Looking in more detail at the data of Figure \ref{fig:fig3}, Ewald-corrected 3-sites models, \textit{i.e.} SPC models (SPC/E and SPC/E\textsubscript{b}) and TIP3P-FB, are rather close to each other, as expected from similar parametrization protocols and representation used. 
Noteworthy, we obtain a similar situation for TIP4P-FB, TIP4P-2005, TIP5P-2018, TIP4P-Ew, and OPC3. This confirms that the number of sites is not the unique critical parameter for obtaining a different representation of water. 

Some interesting observations can be made regarding the 3 more distant outliers of this analysis:
\begin{itemize}
    \item \textit{TIP3P}. Despite being one of the most widely used rigid water models, this appears different from the large cluster of models observed. 
    This result can be interpreted also in light of the limited ability of TIP3P to capture the water solvation shell (Figure \ref{fig:fig2}, right: brown radial distribution function) and its deviation from the experiments in terms of density values (Figure \ref{fig:fig2}, left). 
    \item \textit{TIP4P-ICE}. Here the main discrepancy that can partially explain the $d^{\text{SOAP}}$ from most of the other water models is in the density, which appears to be deviated from the experimental data (Figure \ref{fig:fig2}, left: in light yellow).
    \item \textit{OPC}. This case is interesting because this model is definitely in agreement with all the three observables evaluated in Figure \ref{fig:fig2}. The large $d^{\text{SOAP}}$ distance from the other models presented here could be due to differences in the O-H equilibrium distance ($\sim 0.87$ \AA{ }for OPC) with respect to the others (\textit{e.g.}, $\sim 0.96$ \AA{ }for TIP3P, TIP4P-2005, and TIP4P-Ew, and $1.0$ \AA{ }for SPC/E). This point is corroborated by the comparison of the $g_{OH}(r)$ radial distribution functions, indicating some differences also in the distances of the 1$^{st}$ solvation shell (see Figure S1 in the Supporting Information).
\end{itemize}

Connecting the $d^{\text{SOAP}}$ data of Figure \ref{fig:fig3} to the low dimensional data of Figure \ref{fig:fig2} is nonetheless non-trivial: in fact, the amount of information contained in SOAP data is rich, and strongly related to the local environments that are formed along the MD simulations. 

\begin{figure}[htbp]
\centering\includegraphics[width=0.8\textwidth]{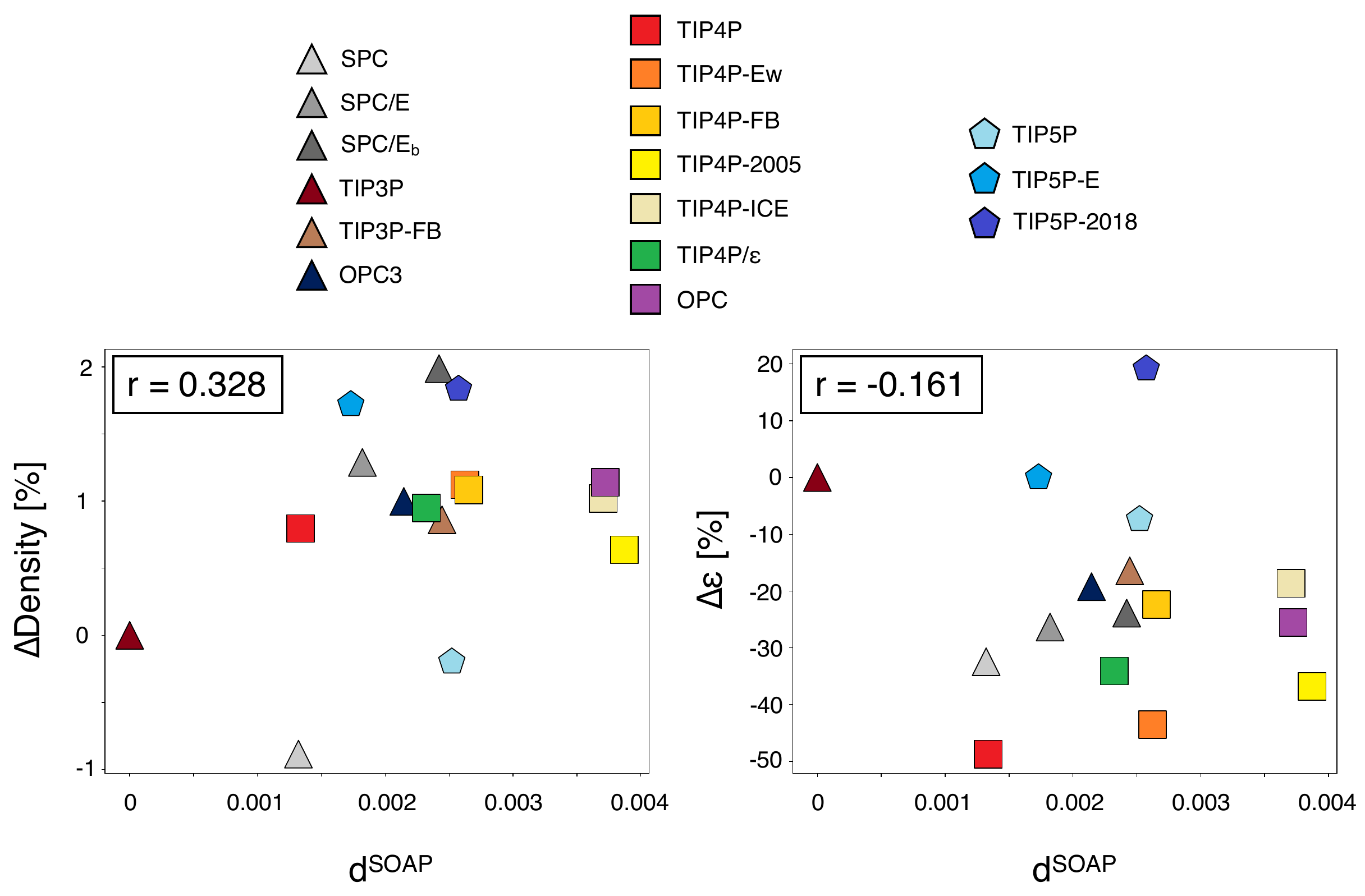}
\caption{Comparison of SOAP distance with density (left) and static dielectric constant (right). We calculated these distances with respect to TIP3P water model; we can observe that the information contained in SOAP power spectra does not correlate neither with density (Pearson correlation coefficient of 0.328), nor with static dielectric constant (Pearson correlation coefficient of -0.161).}
\label{fig:fig4}
\end{figure}

To further confirm the difficulty in this comparison, and the richness of the information contained in SOAP representation, we also computed two scatterplots that compare the $d^{\text{SOAP}}$ between a reference model (here TIP3P) and the relative density and dielectric constant difference (see figure \ref{fig:fig4}).

Figure \ref{fig:fig4} shows that while the density component is an important contributor to the SOAP analysis (and thus SOAP distance), nevertheless it is not enough to completely define and interpret SOAP results. Similarly, also the static dielectric constant does not represent a main contributor to interpret the SOAP distance. 
As a consequence of this, we underline that SOAP is a tool that provides information that could not be obtained via a combination of other macroscopic observables.

\subsection{SOAP analysis of ordered ice phases}

In a recent work, Monserrat \textit{et al.} computed the SOAP spectra of a collection of 1,000 snapshots taken from \textit{ab initio} liquid water simulations.\cite{monserrat2020liquid} Similarly, Engel \textit{et al.} {\cite{engel2018mapping}} studied diverse ice structures employing a combination of SOAP spectra and a dimensionality reduction technique (sketchmap), applied to over 70,000 ice geometries.
Calculating \textit{a posteriori} the SOAP spectra of a set of relevant ice structures, the authors could observe presence of ordered ice-like constructs (\textit{i.e.,} domains in which the arrangement of water molecules is similar to those present in geometrically-optimized ice structures), which can be formed also in liquid water. Such a comparison was performed by using PCA as a dimensionality reduction technique, and by comparing the PCAs of the SOAP spectra obtained for the liquid \textit{vs.} ice aqueous systems. 

Building on such idea, here \textit{we revert the approach}, designing a SOAP-based classification strategy that allows us to automatically recognize ice-like motifs in liquid water MD simulations. 
As a first step, (1) we obtain \textit{ab initio}-optimized reference structures of different ice types; for all these, (2) we compute their characteristic SOAP spectra, and eventually (3) we use these spectra as references for our classification analysis. 

Among the known ice types, in this study we consider a subset of them (investigated also in \cite{monserrat2020liquid}) as the references ice II, ice VIII, ice IX, ice X, ice XIII, ice XIV, ice XV, ice XVII, clathrates sII and sH, ice Ic, ice Ih, and we discard other ones. In particular: 

\begin{itemize}
    \item \textit{Ice Ih}: this is the hydrogen disordered form of hexagonal ice. It accounts for basically all ice naturally found on Earth and thus we include it in our study.;
    \item \textit{Ice III}: this is the hydrogen-bond disordered form of ice IX, and is not considered, while we retain the latter;
    \item \textit{Ice IV}: not considered here, nor in Ref. \cite{monserrat2020liquid} (this ice type displayed issues in the \textit{ab initio} optimization)\cite{monserrat2020liquid};
    \item \textit{Ice V}: this is the hydrogen-bond disordered form of ice XIII, and it is not considered here, while we retain only the latter;
    \item \textit{Ice VI}: this is the hydrogen-bond disordered form of ice XV (reported in ref.\cite{Salzmann2009}) – this is not considered, while we retain only the latter; 
    \item \textit{Ice VII}: this is the hydrogen-bond disordered form of ice VIII, and we retained only the latter;
    \item \textit{Ice X}: displayed issues in the \textit{ab initio} optimization performed in ref.\cite{monserrat2020liquid}, and was not included in that study. However, this ice type is a well-studied ice type, and we include it in our analysis;
    \item \textit{Ice XI}: this is the low-temperature equilibrium structure of hexagonal ice Ih. Since ice Ih is the most common form of ice, here we retain this one, and we discard ice XI;
    \item \textit{Ice XII}: this is the hydrogen-bond disordered form of ice XIV, and we retained only the latter;
    \item \textit{Ice XVI}: this is a low-density clathrate cage topologically equivalent to ice sII. For this reason, we decided to retain only sII and to discard ice XVI in this analysis.
\end{itemize}

\begin{figure}[htbp]
\centering\includegraphics[width=0.75\textwidth]{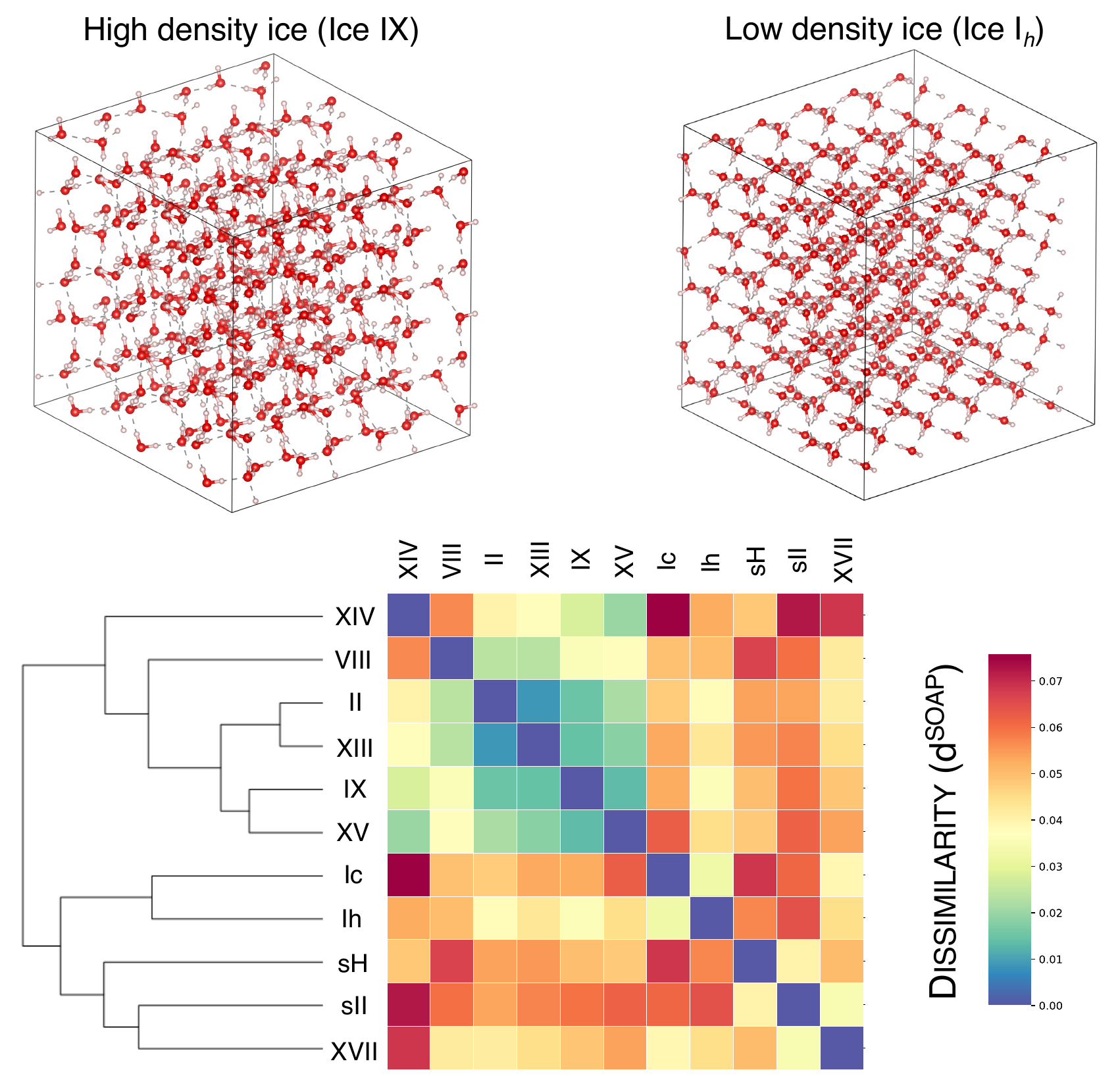}
\caption{SOAP distance matrix between reference ice configurations. Ice X has not been inserted in this matrix for sake of clarity because is far away from all the other ice phases (the complete matrix is available in the Supporting Information, Figure S2).}
\label{fig:fig5}
\end{figure}

In a similar way with respect to the previous section, we obtained the mutual SOAP distance matrix for all these reference ice structures (see Figure \ref{fig:fig5}).  

The most straightforward interpretation of the differences captured in this analysis by the $d^{\text{SOAP}}$ is related to the different ice densities. 
The densest analyzed ice phase, ice X, is found very far from all others (see Figure S2 in the Supporting Information). 
Regarding the other ice types, from this analysis we can identify two main sub-groups (referring to the dendrogram in Figure \ref{fig:fig3}): (i) low-densities ice phases (clathrates ---sII and sH---, Ic, Ih, and XVII) and (ii) high-density phases (II, VIII, IX, XIII, XIV, and XV).

\subsection{Identification of ice-like environments in liquid water}

In the work of Monserrat \textit{et al.}\cite{monserrat2020liquid}, the authors detected the building blocks of different ice phases in MD simulations of liquid water using a machine-learning (ML) potential. This ML potential was trained on quantum calculations at the rev-PBE0\cite{pbe0} level of theory with semiclassical D3 dispersion correction\cite{grimme10} performed on some hundreds of different liquid water conformations.

In particular, the ML potential used by Monserrat et al. allowed to observe in a liquid water system, within a range of 8 \AA{ } around the oxygen atoms of the water molecules, local ice-like environments compatible with those of ice structures.

\begin{figure}[htbp]
\centering\includegraphics[width=0.75\textwidth]{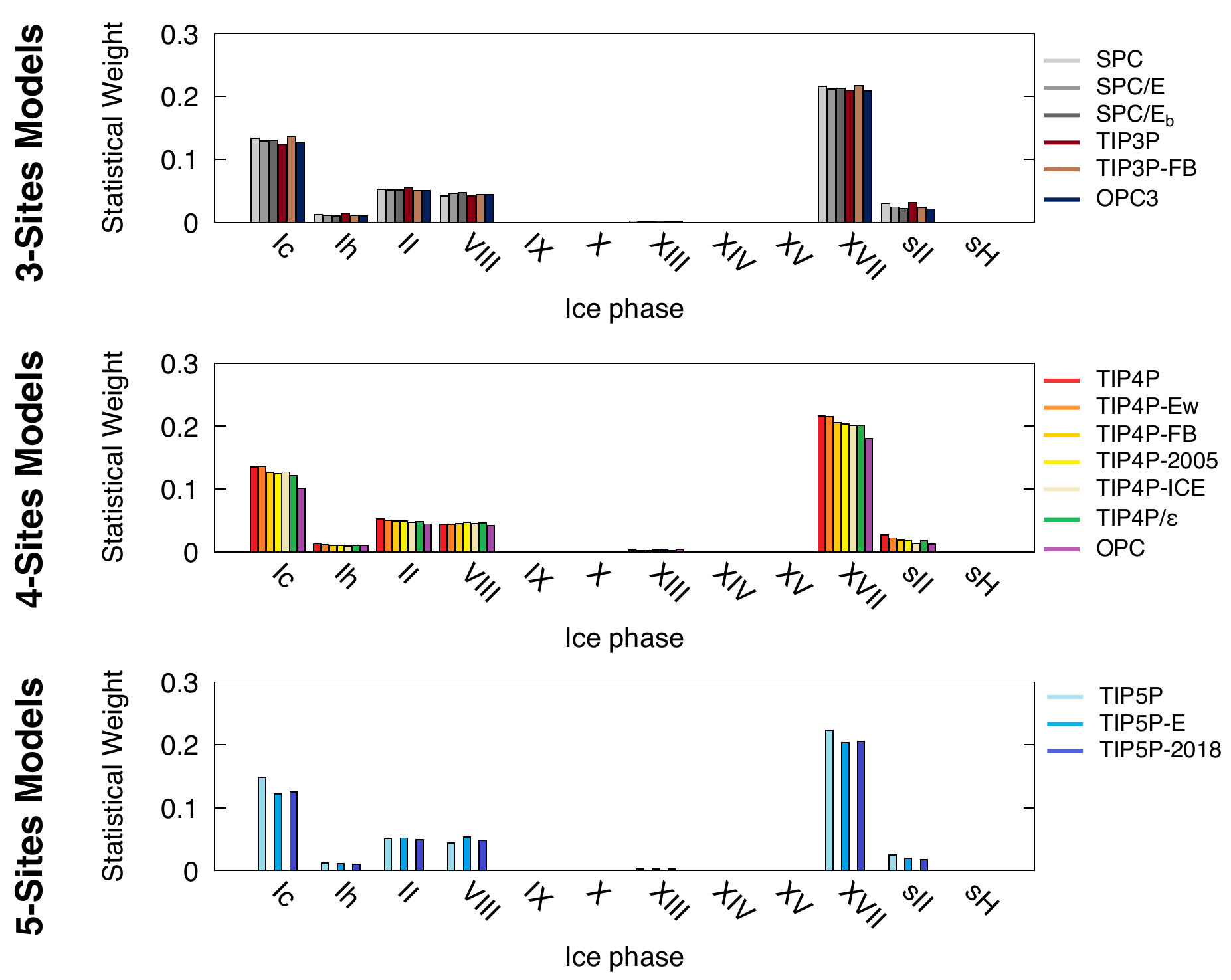}
\caption{Statistics related to ice-like molecular environments in classical MD simulation of liquid water at 298 K. The probability to be in liquid water state was removed for sake of clarity.}
\label{fig:fig6}
\end{figure}

Here we aim to determine to what extent widely used classical rigid water models can capture the same phenomena, using our approach based on the SOAP distance $d^{\text{SOAP}}$ as defined in eq. (\ref{eq:soap_d}) in the Methods section. In particular, we want to label every spectrum (1,024 water molecules for 1,001 MD snapshots, for a total of 1,025,024 SOAP spectra for every simulated model) as closer to a certain reference of either ice or liquid water. 
\\
To achieve this goal, we want to generate reference structure for every state of interest. Operatively (i) we consider the SOAP spectra computed from the geometrically optimized structures of ice considered in the previous section (static ice SOAP references). (ii) Then, to obtain the same reference for liquid water, from every simulation carried out with a specific water model, we extract its average SOAP power spectrum. This is considered as the fingerprint of how each force field describes liquid water (liquid water SOAP reference). 
Finally, for every spectrum obtained from the simulations, (iii) we compute its distance from all the reference SOAP spectra (\textit{i.e.}, average water model \textit{vs.} all the ices), assigning it to the closest one: in other words, all the local environments around oxygen atoms in liquid water simulations are attributed to either (average) liquid water or some ice phases, based on the smallest $d^{\text{SOAP}}$ value.
The results obtained with such analysis are available in Figure \ref{fig:fig6}.

We obtain a statistical weight of $\sim$20\% for water molecules to visit local molecular motifs resembling those proper of ice XVII. 
This is observed consistently in most of the considered classical water models. 
Also Ic, VIII, and II ice-like environments are visited, with a statistically significance of $\sim$10-12\%, $\sim$5\%, and $\sim$4\%, respectively (Figure \ref{fig:fig5}). 
In this analysis, we consider water molecules becoming part of an ordered domain when these are in contact (\textit{i.e.}, when they form hydrogen bonding) with the neighbor ones, forming a network. 
It is worth noting that we never observe large ice domains (crystal nuclei) emerging during the MD runs, as it is expected at room temperature. In fact, our analysis pinpoints the general evidence that the probabilities to find ordered ice-like structures are non-null, and that these can be compared with each other to understand the ordered motifs that emerge along the MD with some statistical relevance.

While sparsely detected along the MD, the formation of other ice-like phases (\textit{e.g.}, ice Ih and sII) is found not significant with respect to the molecular environments reminiscent of ice XVII, Ic, VIII, and II, and of liquid water. As expected for a classical water model,\cite{cheng2019abinitio} standard ice (\textit{i.e.,} ice Ih), conventionally formed at 1 bar pressure and $T<0$\textdegree{}C, is found less favored with respect to ice Ic in liquid water conditions. 

In such a population analysis, we can observe a substantial agreement between all the classical water models analyzed herein (irrespectively to the number of sites, or their parametrization protocols). 
The only (non-substantial) difference detected is for OPC, where the population of Ic and XVII ice-likes visited environments is found smaller than for the other models, while the liquid water population is found larger. This can be imputed to the short O-H intra-molecular distance already discussed above, which favors on average liquid water and reduces the probability to observe the statistical formation of local ordered water motifs having a SOAP spectra similar to that of ices. 
Interestingly, aside from low-density ice-like molecular environments (sII, sH, Ic, Ih, and XVII ice-like domains), we observe that ordered water environments closer to ice VIII, a high-density form of solid water, may also statistically emerge during the MD simulations. 
Summarizing, for all compared classical water models, we observe along the MD runs a non-negligible probability for the emergence of local arrangements of water molecules resembling ices in liquid water at 298 K. 

The presence of local ordered domains that may form in liquid water even far from the freezing temperature is also supported by the experimental evidence. In particular, X-ray spectroscopy revealed a bimodal structural distribution in liquid water\cite{nilsson2015structural}. Furthermore, two distinct peaks in the Raman spectra of liquid water have been also experimentally assigned to molecules involved in ordered (e.g., tetrahedral) H-bond environments vs. a disordered/liquid H-bond network\cite{scherer1974raman}. IR spectra also confirmed similar heterogeneity in liquid water.\cite{marechal2011molecular} These experimental observations find consistency with the SOAP-based results we report herein, providing robustness to our analysis.

\begin{figure}[htbp]
\centering\includegraphics[width=\textwidth]{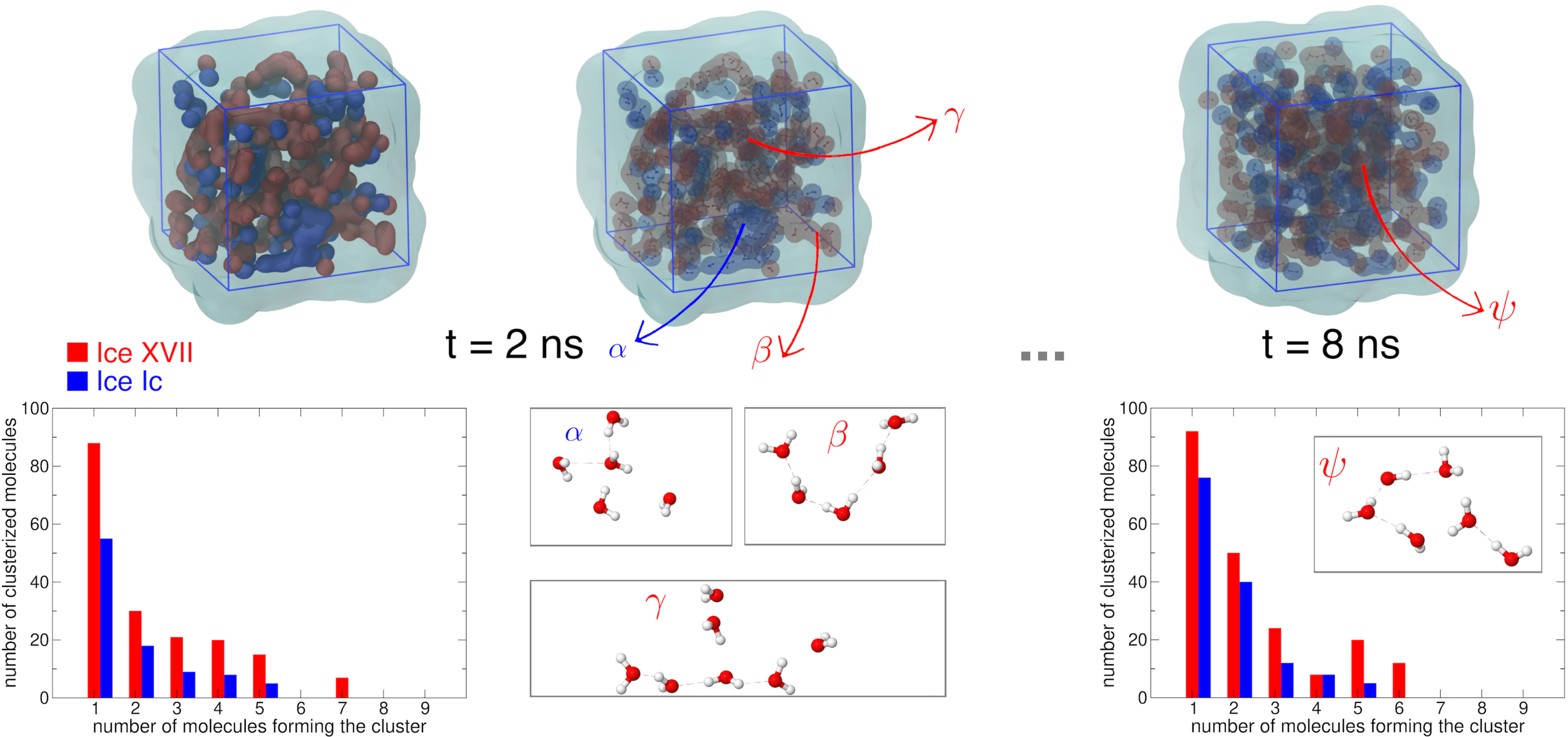}
\caption{Ephemeral ice-like water arrangements visited in liquid water MD simulation. 
Taken from the same MD trajectory of a liquid TIP4P-2005 water box at room temperature, we show 2 MD frames – at 2 ns (left) and 8 ns (right) – separated by a set of dots (...). As it can be observed from the rendered structures (top) and the population histograms (bottom), we do not observe very large ice like-structures along the simulation (we consider 2 molecules in contact if they share an H-bond). 
This is a further hint that the structures observed here at 298 K are in fact metastable and transient, while at lower temperatures they might possibly represent pre-critical ice nuclei.}
\label{fig:fig7}
\end{figure}

To verify how the most representative ice-labeled water molecules are distributed during the MD simulation in the water box, we dissected in more detail a pair of conformations (MD snapshots) obtained from the MD run of TIP4P-2005 water model (Figure \ref{fig:fig7}); for sake of completeness, we specify that in this analysis the only three reference structures considered are the average liquid water simulated with the TIP4P-2005 force field, Ice XVII, and Ice Ic.

Looking more in detail at the individual SOAP obtained for each water molecule in the system at all sampled MD snapshots, we clearly observe that molecules which are labeled within an ice-like environment at a given MD snapshot easily undergo transition in water, or are even observed to become part of a different ice-like local ordered domain at a successive MD snapshot. This demonstrates the ephemeral and statistical nature of such ordered local domains, that are formed and destroyed along the MD, as it is reasonable for a water system of liquid water at room temperature.

We underline the fact that the identification of an ``isolated'' single water molecule classified as ice-like does not mean that it is the only ordered molecule (no order can exist with just one molecule). In fact, in our SOAP analysis, the cutoff threshold $r_{cut}$ is equal to 1 nm, which accounts for more than the first 3 solvation shells around each water molecule. This means that even in the case of clusters composed of one single molecule, we have a water molecule in the simulation box around which, within a spherical volume of radius 1 nm, the neighbor water molecules are found distributed in a way that is closer to a specific (ordered) ice type rather than to the average (disordered) distribution typical of liquid water. The analysis of the lifetime of such ice-like water environments indicated that this is on average very short, namely, in the order of $\sim$ 10 ps or below. This is reasonable, as we are at room condition, where the formation of ordered ice-like clusters is highly unfavored, and it underlines the ephemeral nature of these pseudo-ordered environments in such conditions.

Summarizing, the strength of this analysis relies in the fact that is fully data-driven, and it is based on a general and abstract high-dimensional descriptor of the molecular environments that surround each water molecule in the system. The SOAP environments that are sampled during the MD simulation are then compared with ice-environments/configurations contained in a reference dataset. It is worth noting that this analysis is quite flexible and abstract, and rather unbiased - namely, it is intrinsically limited only by the set of ice structures used as reference points in the dataset, and by the time and space resolution used for the calculation of the SOAP spectra (i.e., the cutoff and the timestep between the sampled snapshots). Furthermore, it can be in principle applied to assess similarities/differences in the local molecular configurations of any system (even different from water) for which well-characterized reference configurations are available.

\section{Conclusions}
In this work, we compare how classical water models represent liquid water macroscopically and microscopically, using a metrics that is built on a high-dimensional SOAP framework. The general behavior of rigid water models analyzed here is found overall similar, but some differences in the studied observables can still be detected (as also shown in an extensive way by the recent work of Kadaoluwa Pathirannahalage \textit{et al.}\cite{pathirannahalage2021systematic}).
With our approach we easily detect models with considerable differences in the monitored observables (mainly TIP3P and TIP4P-ICE), but also another state-of-the-art model (OPC) that displays a very peculiar shorter H-O intramolecular bond, and thus is correctly identified as an outlier as well. 

We compare by means of our SOAP distance a collection of ice structures obtained from \textit{ab initio} geometrical refinement of experimental ones. 
We confirm that our SOAP analysis mainly separates ice structures on the basis of their packing (and thus density), like already shown by Monserrat \textit{et al.}\cite{monserrat2020liquid}. 
Comparing the SOAP distances obtained in this analysis, we can qualitatively observe that the differences between ices are significantly larger (1 order of magnitude) than those which can be observed between the various studied liquid water models. This is reasonable because in all the simulations of the liquid phase transient domains resembling ices appear. Thus, said models share a large number of possible structural realizations and their close $d^{\text{SOAP}}$ between averaged spectra reflects this. On the contrary, SOAP distances on the static structures of ices display the inherent differences among their geometries.

Combining our SOAP distance, with a classification of disordered model systems (liquid water) based on ordered reference systems (ice structures), we could clearly identify the formation of ice-like local molecular environment. In other words, water molecules arrange with respect to each other in an ordered configuration reminiscent of that proper of molecules in different ice types, even in liquid water at room conditions. 
Computing the SOAP distances between every water molecule in the system at each MD simulation step \textit{vs.} reference ice structures and average water displacement, we could quantify the relative probabilities to observe the emergence of different types of local ordered domains in the liquid systems along the equilibrium MD. 
We find a non-negligible probability for having water molecules becoming part of ordered local domains, more similar to ice-like elementary domains than to liquid water, using all the classical water force fields analyzed. 
This observation is in agreement with recent investigations performed on \textit{ab initio} or \textit{ab initio}-trained Machine Learning potentials\cite{monserrat2020liquid,cheng2020mapping}, as well as in high-dimensional analysis of classical water\cite{offei2021high}. On this last point, this observation is valid in \textit{both} quantum and classical simulations of water, and thus it is reasonable to wonder if this is an \textit{artefact} due to the very way that water is currently modeled by computational means, or if it is an actual phenomenon which is correctly modeled and is still unnoticed experimentally. \\
We also observe the characteristic ephemeral and statistical nature of such ordered water environments, from which water molecules can exchange in an out, and which may dynamically emerge and disappear along the MD simulations, as it can be expected for a liquid water system. 
The high-dimensional analysis reported herein can be extremely useful to characterize complex molecular systems where ordered and disordered states coexist in dynamical exchange and equilibrium. \\
Such a classification approach holds a considerable potential for the development of more accurate and reliable force fields at an affordable computational cost, as well as to study and classify a variety of self-organizing systems, not necessarily restricted to water models. \\
As a last point, the fact that (ephemeral) local ordered domains arise even in liquid water, and quite far from the freezing temperature, finds consistency with the available experimental evidence (e.g., X-ray, Raman, IR spectroscopy)\cite{nilsson2015structural,scherer1974raman,marechal2011molecular,camisasca2019proposal,walrafen1997}. On the one hand, this supports the results reported herein. On the other hand, this underlines the importance of identifying such ephemeral ordered domains, and studying them, as these may play a central role in global thermodynamic properties of water, as well as phase transitions and ice nucleation processes upon temperature decrease.

Summarizing, the outcomes of this work are interesting from multiple point of view. 
First, our results demonstrate that liquid-like ordered water environments may emerge also in relatively simple rigid water models. Such models are widely used to simulate a variety of systems in water. This makes these observations relevant in multiple directions, considered, for example, that it has been demonstrated recently how liquid water can play a key role in controlling host-guest interactions{\cite{rizzi2021role}}, and how such ephemeral ordered domains may play an important role in, e.g., the processes of ice nucleation in liquid water during freezing.

\section{Methods}

\subsection{MD Simulations}
We created a cubic box containing 1,024 water molecules, arranged in initial random configurations, using Packmol\cite{martinez2009packmol}. All simulations were conducted using GROMACS version 2021.2\cite{abraham2015gromacs}. All model systems were first minimized using a steepest descent algorithm for $2\cdot10^{3}$ steps, and then equilibrated \textit{via} 1 ns of MD simulation conducted in $NVT$ conditions (with the constants $NVT$ being the number of particles, the simulation volume, and the simulation temperature, respectively). 
As a last preparation step, we equilibrated the volume of the simulation box \textit{via} 1 ns of MD simulation in $NpT$ conditions (with the constants $NpT$ being the number of particles, the  pressure, and the simulation temperature, respectively). 
All MD runs used a 2 fs timestep. 
After these equilibration phases, we ran 10 ns long production MD runs with the same conditions as the last equilibration phase (NPT). 
For all those steps, we kept temperature with velocity rescale thermostat\cite{bussi2007canonical} (with coupling constant of 0.2 ps) and the pressure was maintained constant (1 bar) with an isotropic scheme using the cell rescale barostat\cite{bernetti2020pressure} (with a coupling constant of 2 ps, and a compressibility of $4.5\cdot 10^{-5}$ bar\textsuperscript{-1}). 
A cutoff distance of 1 nm was used for short-range electrostatic and van der Waals interactions, and the long-range interactions were computed with the particle-mesh Ewald summation method\cite{essmann1995smooth}. Long-range dispersion corrections to the pressure and potential energy were considered\cite{shirts2007accurate}. 
We pinpoint that this choice of PME for long-range interactions can cause some deviation from the original values in the oldest water models in their original simulation conditions. 
This choice has been made nevertheless to have the best common ground for a fair comparison between different classical water models used herein (and to employ simulation protocols as similar as possible to the ones routinely used nowadays).

For all the water models, we performed a simulation from 273 K to 373 K (with a temperature-step of 10 K), and a last block of simulations performed at 298 K (25 \textdegree{}C) from which we computed radial distribution functions for O-O (see Figure \ref{fig:fig3}) and for O-H (see Figure S1 in Supporting Information).

\subsection{Observables}
During our analysis we computed a series of observables, namely liquid water density ($\rho_{L}$) as a function of temperature, static dielectric constant ($\varepsilon$) as a function of temperature, and the radial distribution function between oxygen atoms ($g_{OO}$) at 298 K.

\textbf{Liquid water density.} We computed the mass density of the water models by calculating the size of the simulation box, namely as:
\begin{equation}
    \rho_{L} = \frac{N \cdot m_{H_{2}O}}{\mathcal{N}_{A} \cdot V_{\text{box}}} ~,
\end{equation}
where $N$ is the number of water molecules (1024 in our case), $m_{H_{2}O}$ is the mass of the water molecules in a.u., $\mathcal{N}_{A}$ is the Avogadro's number and $V_{\text{box}}$ is the simulation box volume. This analysis was performed with PLUMED plugin version 2.7.\cite{tribello2014plumed,bonomi2019promoting}

\textbf{Static dielectric constant.} We computed the static dielectric constant from the fluctuations of the total dipole moment $M$ of the simulation box, namely as:
\begin{equation}
    \varepsilon = 1 + \frac{\langle M^{2} \rangle - \langle M \rangle^{2}}{3 \varepsilon_{0}Vk_{B}T} ~,
\end{equation}
where $\varepsilon_{0}$ is the vacuum permittivity, $V$ the volume of the simulation box, $k_{B}$ the Boltzmann constant, $T$ the system temperature, and $\langle \cdot \rangle$ represents the thermodynamic average. 
We computed this observable by using the \texttt{gmx dipoles} routine in the GROMACS\cite{abraham2015gromacs} suite.

\subsection{SOAP}
Smooth Overlap of Atomic Position (SOAP)\cite{bartok2013,de2016comparing} is an atomic environment descriptor that aims to translate the 3D configuration of a set of atoms around a center (that can be an atom, or more in general a point in space) to a high-dimensional representation. The result of such analysis is a vector called ``power spectrum''. \\
Given a system conformation $\Gamma$ in the 3D space, the SOAP power spectrum calculation is done by expanding the local atomic/particle density $\rho_i(\Gamma,\vec{r})$ (defined in the neighborhood of every SOAP center within a spatial cutoff, $r_{cut}$) projecting it onto a basis of orthogonal radial functions ${g_n(r)}$ and spherical harmonics $Y_{lm}(\theta,\phi)$, which for the \textit{i}-th site can be expressed as:
\begin{equation}
\label{eq:density}
    \rho_i(\Gamma,\vec{r}) = \sum_{j\in r_{cut}} \sum_{nlm} c^j_{nlm}(\Gamma) g_n(r) Y_{lm}(\theta,\phi) ~,
\end{equation}
where the $j$ index runs over all the sites in the cutoff. In our case, one SOAP is centered on the oxygen atom of each water molecule in the simulated systems. At each sampled MD snapshot, we thus obtain 1,024 SOAP spectra, characteristic for the environment that surrounds each individual water molecule in the system. The analysis is then repeated for 1,001 frames obtained from the equilibrium MD simulations, obtaining a total of 1,025,024 individual SOAP spectra for all simulated water models.\\
It is worth noting that $\rho_i(\Gamma,\vec{r})$ is multi component (it has one component for each chemical species). 
From the eq. (\ref{eq:density}) we can obtain the SOAP power spectrum vector defined as
\begin{equation}
    \label{eq:powerspectrum}
    \vec{p}(\Gamma)_{nn'l} = \pi \sqrt{\frac{8}{2l+1}} \sum_{m=-l}^l c^*_{nlm}(\Gamma)c_{n'lm}(\Gamma) ~,
\end{equation}
which encodes all the information of the atomic environment.
Eq. (\ref{eq:powerspectrum}) represents also the computational output obtained from the SOAP calculation using the DScribe\cite{dscribe} package.

A similarity measure between two environments centered in two sites can be formally defined by building a linear kernel of their density representations. Such kernel can be reduced to the dot product of the two sites' SOAP power spectra as:\cite{bartok2013}
\begin{equation}
    \label{eq:soap_k}
    K^{SOAP}(i,j) = \rho_i(\Gamma,\vec{r}) \cdot \rho_j(\Gamma,\vec{r}) \propto \vec{p}_i \cdot \vec{p}_j ~.
\end{equation}
Eq. (\ref{eq:soap_k}) can be interpreted as a measure of how much the two local environments (surrounding a water molecule) are superimposed to each other (\textit{i.e.}, how similar they are). The value of $K^{SOAP}$ ranges from 0 for completely different to 1 for exactly matching environments.

From eq.(\ref{eq:soap_k}), we can further define a metrics (that we call ``SOAP distance'') between two environments, namely as:
\begin{equation}
    \label{eq:soap_d}
    d^{\text{SOAP}}(i,j) = \sqrt{2 - 2 \cdot K^{SOAP}(i,j)} \propto \sqrt{2 - 2 \vec{p}_i \cdot \vec{p}_j} ~,
\end{equation}
where $\vec{p}_i$ is the \textit{i}-th center's power spectrum.
Both SOAP kernel and distance representations give a bounded measure of how similar two local sites/environments are (\textit{i.e.}, how their local densities are superimposable in the power spectra space).

\begin{acknowledgements}
We thank Andrea Gardin and Carlo Camilloni for useful discussion, and Daniele Rapetti for reviewing the code used in this work. GMP acknowledges the funding received by the European Research Council (ERC) under the European Union's Horizon 2020 research and innovation programme (grant agreement no. 818776 - DYNAPOL). 
\end{acknowledgements}

\section*{Data Availability}
All the input files and relevant analysis script are available on GitHub at the address  \url{https://github.com/GMPavanLab/Water_SOAP}.

\section*{Supporting Information}
Additional figures on our analyses (O-H radial distribution function, full SOAP distance matrix for ice and distribution of $d^\text{SOAP}$) are available in the Supporting Information PDF file.

\bibliography{biblio}

\end{document}


\maketitle


\begin{figure}
\centering\includegraphics[trim={0cm 0cm 0cm 0cm},clip,width=\textwidth]{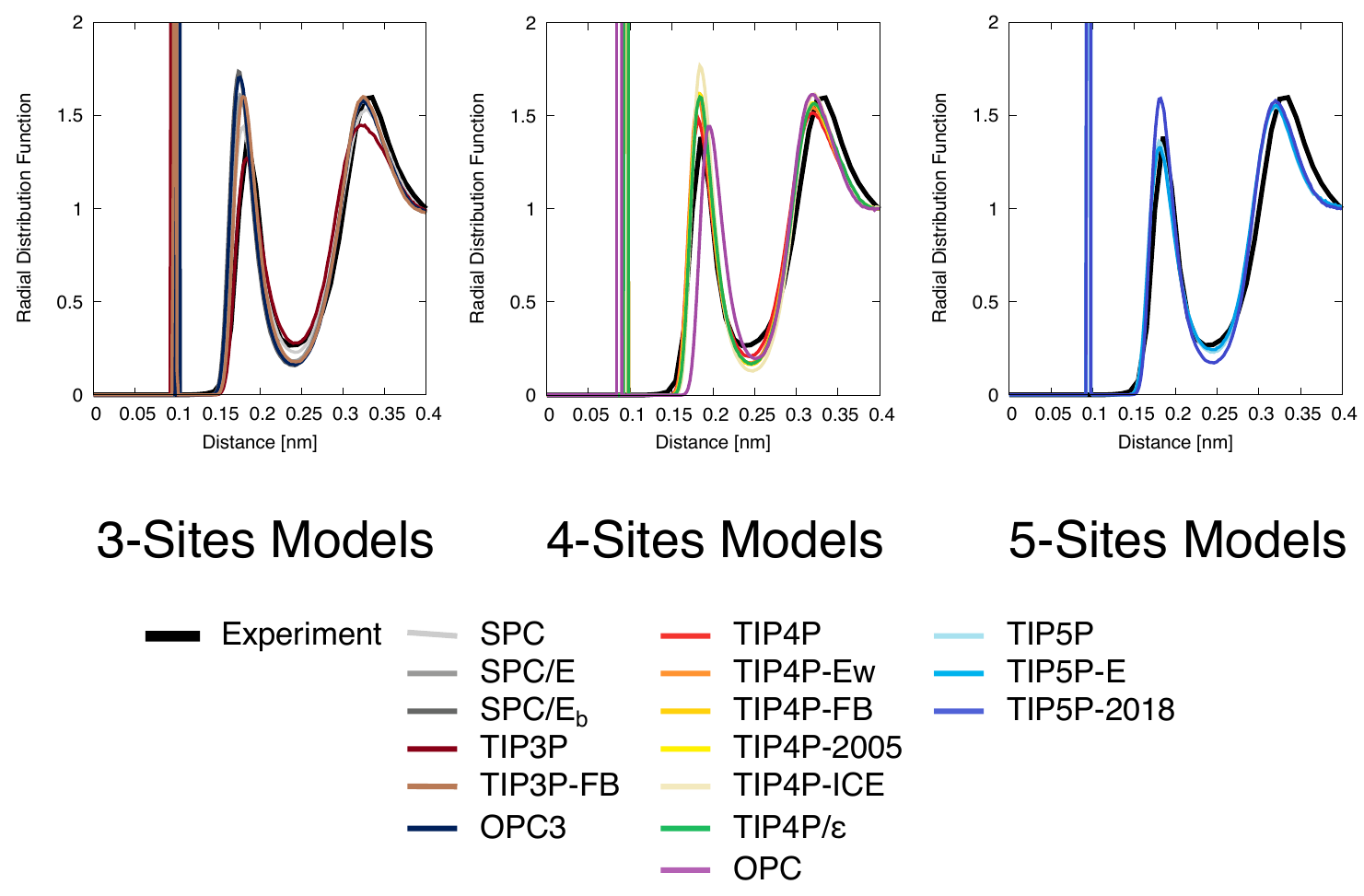}
\caption{Radial distribution functions between Oxygen and Hydrogen atoms in different water models.}
\label{fig:rdf_OH}
\end{figure}

\begin{figure}
\centering\includegraphics[trim={0cm 0cm 0cm 0cm},clip,width=\textwidth]{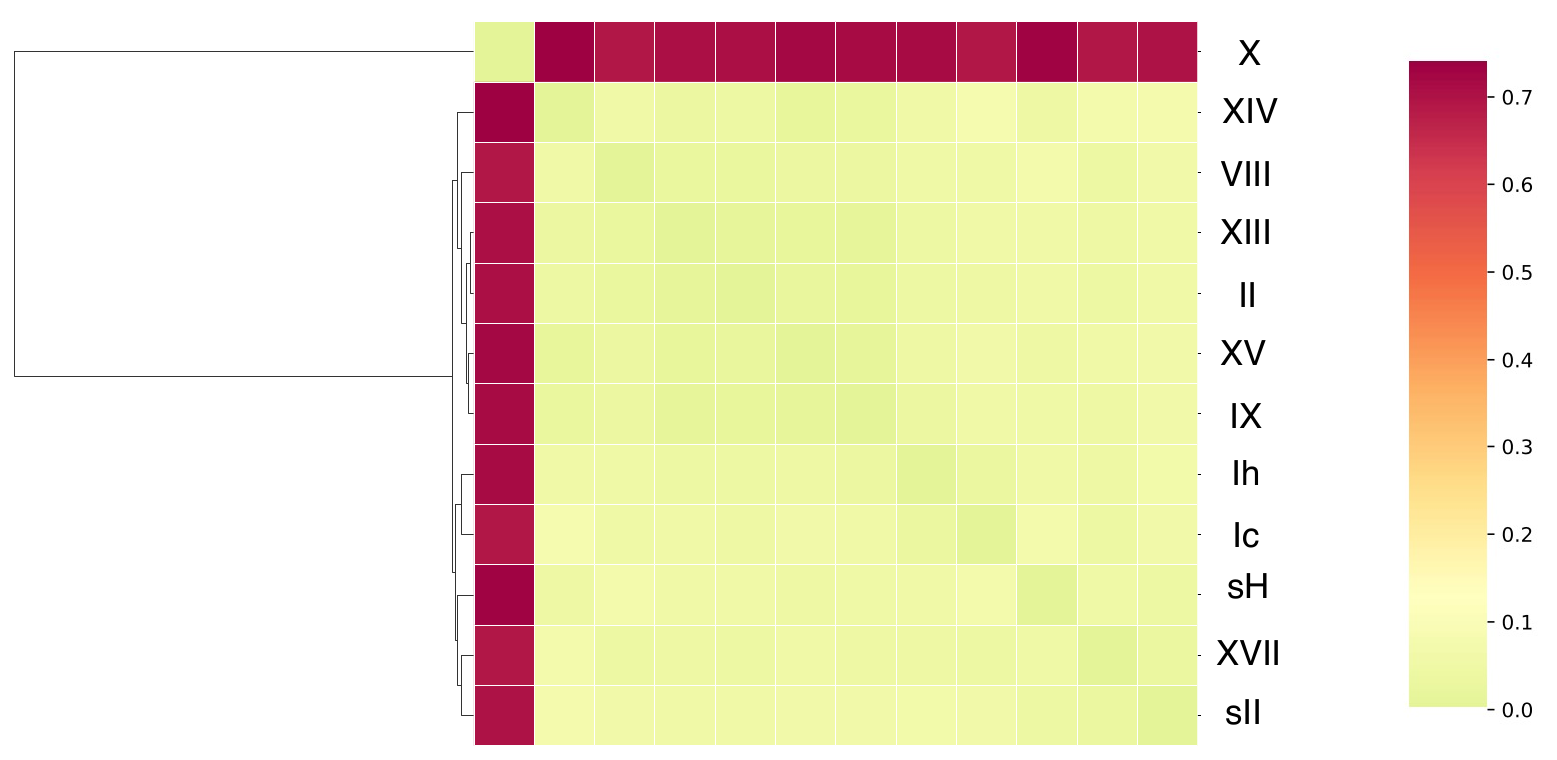}
\caption{SOAP distance matrix between reference ice configurations, including ice X.}\label{fig:dist_mat_x}
\end{figure}

\begin{figure}
\centering\includegraphics[trim={0cm 0cm 0cm 0cm},clip,width=\textwidth]{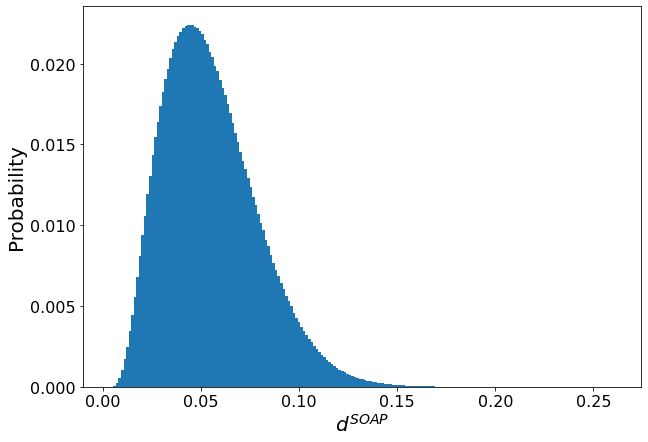}
\caption{Distribution of $d^{\text{SOAP}}$ among different environments from TIP3P simulation at 298 K.}\label{fig:distr_dsoap}
\end{figure}
